\documentclass{llncs}

\usepackage{amssymb}
\usepackage{amsmath}

\usepackage{xcolor,colortbl}
\usepackage{graphicx}

\usepackage{algorithm}
\usepackage{algorithmic}

\usepackage{url}

\usepackage{float}
\usepackage{booktabs}

\usepackage{tikz}
\usetikzlibrary{calc,shapes}

\usepackage{framed} 

\usepackage{aicescover}

\usepackage{tabularx}

\newcolumntype{I}{!{\vrule width 1.3pt}}
\newlength\savedwidth

\newcommand{\myFlaTwoByTwo}[4]{
\renewcommand{\arraystretch}{1.2}
\setlength{\arraycolsep}{0pt}
  \left(
	\begin{array}{c@{\;\;}|@{\;\;}c} 
      #1 & #2 \\ \hline
      #3 & #4 
    \end{array} 
  \right)
}

\newcommand{\myFlaTwoByOne}[2]{
\renewcommand{\arraystretch}{1.2}
  \left(
	\begin{array}{c} 
      #1 \\ \hline
      #2  
    \end{array} 
  \right)
}

\newcommand{\myFlaOneByTwo}[2]{
\renewcommand{\arraystretch}{1.2}
  \left(
	\begin{array}{c@{\;\;}|@{\;\;}c} 
      #1 & #2
    \end{array} 
  \right)
}

\newlength{\tabwidth}
\newlength{\tabmargin}
\newcommand{\omicols}[0]{{\sc ols-grid}}
\newcommand{\qr}[0]{{\sc qr}}
\newcommand{\gemm}[0]{{\sc gemm}}
\newcommand{\gemv}[0]{{\sc gemv}}
\newcommand{\trsm}[0]{{\sc trsm}}
\newcommand{\trsv}[0]{{\sc trsv}}

\colorlet{mtblas}{green!60!black}
\colorlet{openmp}{blue!60!black}
\colorlet{io}{red!60!black}

\begin{document}

\title{Large-scale linear regression: Development of high-performance routines}

\author{Alvaro Frank \and Diego Fabregat-Traver \and Paolo Bientinesi}

\institute{
    AICES, RWTH Aachen, 52062 Aachen, Germany \\[1mm]
  \email{alvaro.frank@rwth-aachen.de, \\
         fabregat@aices.rwth-aachen.de, \\
         pauldj@aices.rwth-aachen.de}
}

\aicescovertitle{Large-scale linear regression: Development of high-performance routines}
\aicescoverauthor{Alvaro Frank, Diego Fabregat-Traver and Paolo Bientinesi} 
\aicescoverpage

\maketitle

\begin{abstract}

In statistics, series of ordinary least squares problems (OLS) are used to
study the linear correlation among sets of variables of interest; in many
studies, the number of such variables is at least in the millions, and the
corresponding datasets occupy terabytes of disk space.  As the availability of
large-scale datasets increases regularly, so does the challenge in dealing with
them. Indeed, traditional solvers---which rely on the use of ``black-box''
routines optimized for one single OLS---are highly inefficient and fail to
provide a viable solution for big-data analyses.  As a case study, in this
paper we consider a linear regression consisting of two-dimensional grids of
related OLS problems that arise in the context of genome-wide association
analyses, and give a careful walkthrough for the development of {\sc ols-grid},
a high-performance routine for shared-memory architectures; analogous steps are
relevant for tailoring OLS solvers to other applications.  In particular, we
first illustrate the design of efficient algorithms that exploit the structure
of the OLS problems and eliminate redundant computations; then, we show how to
effectively deal with datasets that do not fit in main memory; finally, we
discuss how to cast the computation in terms of efficient kernels and how to
achieve scalability. Importantly, each design decision along the way is
justified by simple performance models.  {\sc ols-grid} enables the solution of
$10^{11}$ correlated OLS problems operating on terabytes of data in a matter of
hours. 

\vspace{2mm}
\noindent
{\bf  Keywords:}
    Linear regression, 
    ordinary least squares, 
    grids of problems,
    genome wide association analysis,
    algorithm design,
    out-of-core,
    parallelism,
    scalability
\end{abstract}

\section{Introduction}
\label{sec:intro}

Linear regression is an extremely common statistical tool for modeling the
relationship between two sets of data. 
Specifically, given
a set of ``independent variables'' $x_1, x_2, \dots, x_p$, and a
``dependent variable'' $y$, one seeks the 
correlation terms $\beta_i, i=1, \dots, p$ 
in the linear model 
\begin{equation}
  \label{eq:linreg}
  y = \beta_1 x_1 + \dots \beta_p x_p.
\end{equation}
In matrix form, Eq.~\eqref{eq:linreg} is expressed as 
$ \mathbf{y} = X \mathbf{\beta} + \mathbf{\epsilon}$, 
where $ \mathbf{y} \in R^n$ is a vector of $n$ ``observations'', 
the columns of $X \in R^{n \times p}$ are ``predictors'' or ``covariates'', 
the vector $\mathbf{\beta} = [\beta_1,\dots,\beta_p]^T$
contains the ``regression coefficients'', 
and $\mathbf{\epsilon}$ is an error term that one wishes to minimize.
In many disciplines, linear regression is used to quantify the relationship
between one or more $\mathbf{y}$'s from the set $\mathcal{Y}$,
and each of many $x$'s from the set $\mathcal{X}$.
The computational challenges raise from the all-to-all nature of the
problem (estimate how strongly each of the covariates is related to each of the observations), and
from the sheer size of the datasets $\mathcal{Y}$ and $\mathcal{X}$, 
which often cannot be stored directly in main memory.

One of the standard approaches to fit the model~\eqref{eq:linreg} to given
$\mathbf{y}$ and $X$ is by solving an ordinary least squares (OLS) problem; 
in linear algebra
terms, this corresponds to computing the vector $\mathbf{\beta}$ such that
\[
\mathbf{\beta} = \left( X^T X \right)^{-1} X^T \mathbf{y}.
\]
In typical datasets, $\hat m$, the number of available covariates ($\hat m =
|\mathcal{X}|$), is much larger than $p$, the number of variables actually used
in the model. 
In this case, a group of $l<p$ covariates is kept fixed, 
and the remaining $p-l$ slots are filled from $\mathcal{X}$, in a rotating fashion;
it is not uncommon that the value $p-l$ is very small, often just one, thus originating 
$m \ge \hat m$ distinct OLS problems. 
Mathematically, this means computing a series of $\beta_i$'s such that
\begin{equation}
  \label{eq:1dgrid}
  \mathbf{\beta}_i = \left( X_i^T X_i \right)^{-1} X_i^T \mathbf{y}, \quad
  \text{where}\quad i=1,\dots,m;  
\end{equation}
here $X_i$ consists of two parts: $X_L$, which contains $l$ columns and is
fixed across all $m$ OLS problems, and $X_{R_i}$, which instead contains $p-l$
columns taken from $\mathcal{X}$. In many applications, $m$ can be of the
order of millions or even more.

When $t>1$ dependent variables ($t = |\mathcal{Y}|$) are to be studied against
$\mathcal{X}$, the problem~\eqref{eq:1dgrid} assumes the more general form
\begin{equation}
  \label{eq:2dgrid}
  {\beta}_{ij} = \left( X_i^T X_i \right)^{-1} X_i^T \mathbf{y}_j, \quad
\end{equation}
where $i=1,\dots,m$, and $j=1,\dots,t$,   
indicating that one has to compute a two-dimensional grid of 
${\beta}_{ij}$'s, each one corresponding to an OLS problem.
This is for instance the case in genomics (multi-trait genome-wide
  association analyses)~\cite{Hindorff-2009} 
  and econometrics (explanatory variable
 exploration)~\cite{Sala97}.

Despite the fact that OLS solvers are provided by many
libraries and languages (e.g., LAPACK, NAG, MKL, Matlab, R),
no matter how optimized those are, 
any approach that aims at computing the 2D grid~\eqref{eq:2dgrid} 
via $t \times m$ invocations of a ``black-box'' routine 
is entirely unfeasible.
The main limitations come from the fact that this approach 
leads to the execution of inefficient and redundant operations, 
lacks a mechanism to effectively manage data transfers from and to hard disk, and
underutilizes the resources on parallel architectures.

In this paper, we consider an instance of Eq.~\eqref{eq:2dgrid} as it arises
in genomics, and develop {\sc ols-grid}, a parallel solver tailored for this application.
Specifically, we focus on the study of {\em omics} data\footnote{With the term
{\em omics} we refer to large-scale analyses involving at least hundreds
of traits~\cite{gieger08,Demirkan2012,OmicABEL}.}
in the context of genome wide association analyses (GWAA).%
\footnote{GWAA are often also referred to as genome wide association studies (GWAS)
and whole genome association studies (WGAS).}
Omics GWAA study the relation between $m$ groups of genetic markers and $t$
phenotypic traits in populations of $n$ individuals. In terms of OLS, each
trait is represented by a vector $y_j$ containing the trait measurements (one
per individual); each matrix $X_i = [X_L | X_{R_i}]$ is composed of a set of
$l$ fixed covariates such as sex, age, and height ($X_L$), and one of the
groups of $r = p - l$ markers $(X_{R_i})$.  A positive correlation between
markers $X_{R_i}$ and trait $y_j$ indicates that the markers may have an
impact in the expression of the trait.

Typical problem sizes in omics GWAA are roughly
$10^3 \le n \le 10^5$,
$2 \le p \le 20$ (with $r=1$ or $2$),
$10^6 \le m \le 10^8$, and
$10^2 \le t \le 10^5$.
An exemplary analysis with size 
$n = 30{,}000$,
$p = 10$ ($l=8$, $r=2$),
$m = 10^7$, and
$t = 10^4$,
poses three considerable challenges.
First, it requires the computation of $10^{11}$ OLS problems, which, if tackled by
a traditional ``black-box'' solver, would perform O($10^{18}$) floating
point operations (flops).
Despite the fact that the problem lends itself to a lower computational cost
and efficient solutions, a black-box solver ignores the structure of the
problem and requires large clusters to obtain a solution.
The second challenge is posed by the size of the datasets to be processed: 
assuming single-precision data (4 bytes per element), a GWAA solver
reads as input about 2.4 TBs of data and  produces as output 4 TBs of data.
If the data movement is not analyzed properly, 
the time spent in I/O transfers might render the computation unfeasible.
Finally, the computation needs to be parallelized and organized 
so that the potential of the
current multi-core and many-core architectures
is fully exploited.

\paragraph{Contributions.}
This paper is concerned with the design and the implementation of
{\sc ols-grid}, a
high-performance algorithm for large-scale linear regression.
While we use omics GWAA as a case study, the discussion is
relevant to a range of OLS-based applications.
Specifically, we
1) illustrate how to take advantage of the specific structure in the grid
of OLS problems to design specialized algorithms,
2) analyze the data transfers from and to disk to effectively deal 
with large datasets that do not fit in main memory, and
3) discuss how to cast the computation in terms of efficient
kernels and how to attain scalability on many-core architectures.
Moreover, by making use of simple performance models, we identify the
performance bottlenecks with respect to the problem size. 
{\sc ols-grid}, available as part of the GenABEL suite~\cite{genabel},
allows one to execute an analysis of the aforementioned size in less than 7 hours
on a 40-core node.

\paragraph{Related work.}
Genome-wide association analyses received a lot of attention in the last
decade~\cite{GWAScatalog}. Numerous high-impact findings have been reported,
including but not limited to the identification of genetic variations
associated to
a common form of blindness,
type 2 diabetes, and
Parkinson's disease~\cite{Hageman17052005,Edwards15042005,Frayling2007,Nalls2014}.
A popular approach to GWAA is the so called Variance Components model,
which boils down to a set of equations similar to Eq.~\eqref{eq:2dgrid}.
The main difference with the present work lies on the core equation, 
where one has to solve grids of generalized least squares (GLS) problems instead of grids of OLSs.
A number of libraries have been developed to carry out GLS-based GWAA,
the most relevant being FaST-LMM, GEMMA, GWFGLS, and OmicABEL~\cite{genabel,Lippert2011,Zhou2012,OmicABEL}. 

OmicABEL, developed within our research group, showed remarkable
performance improvements with respect to the other existing methods~\cite{SingleGWAS,MultiGWAS}.
Clearly, the same library can be used, by setting the covariance matrix to
the identity, to solve Eq.~\eqref{eq:2dgrid}. While possible, 
this is not advisable: OmicABEL reduces the two-dimensional grid 
of GLS problems to the grid of OLS problems 
$$ \tilde{b}_{ij} = 
(\tilde{X}_{ij}^T \tilde{X}_{ij})^{-1} \tilde{X}_{ij}^T \tilde{y}_{j}, $$
which is deceivingly similar to Eq.~\eqref{eq:2dgrid};
however, the subtle differences in the
dependencies (subindices) of the design matrix $X$ 
lead to a more expensive and less efficient algorithm. 
In Sec.~\ref{sec:experiments} we show how the new 
{\sc ols-grid} outperforms OmicABEL-Eig considerably.

A number of tools are focused on OLS-based linear regression analyses for
GWAA; among them, we mention 
ProbABEL (also part of the GenABEL suite), GWASP, and
BOSS~\cite{Aulchenko-Struchalin-2010,gwasp,BOSS}.
GWASP stands out for its elegant algorithmic approach and a
performance-oriented design; a more detailed discussion is given in 
Sec.~\ref{sec:comparison}.

\paragraph{Organization of the paper.}

The remainder of the paper is organized as follows.  In
Sec.~\ref{sec:algorithm} we describe the design of an efficient algorithm that
exploits problem-specific knowledge. In Sec.~\ref{sec:ooc} we analyze the data
transfers and discuss possible limitations inherent to the problem at hand,
while in Sec.~\ref{sec:hp} we focus on the high performance and scalability of
our software.  Section~\ref{sec:experiments} presents performance results.
Finally, Sec.~\ref{sec:conclusions} draws conclusions and discusses future
work.

\section{From the problem specification to an efficient algorithm}
\label{sec:algorithm}

\sloppypar
In this section we discuss the steps leading to the core algorithm behind {\sc ols-grid}.
Starting from an elementary and generic algorithm, 
we incrementally refine it into an efficient algorithm 
tailored specifically for grids of OLS arising in GWAA studies. 
The focus is on the reduction of the computational complexity.

As a starting point to solve Eq.~\eqref{eq:2dgrid}, 
we consider a most naive black-box solver consisting of
two nested loops (for each $i$ and $j$) around 
a QR-based algorithm for
OLS~\cite{Golub:1996:MC:248979}:
$$
\begin{aligned}
    \{Q, R\} & := qr(X) \\
    b & := R^{-1} Q^T y.
\end{aligned}
$$
Since the design matrix $X$ only depends on the index $i$, the loop ordering $i, j$
makes it possible to compute the QR factorization once and reuse it across the 
$j$ loop. Even with this simple code motion optimization,
this first algorithm performs substantial redundant calculations
due to the structure of $X_i$.

Recall that each matrix $X_i$ can be logically seen as consisting
of two parts $(X_L | X_{R_i})$, where $X_L$ is constant,
and $X_{R_i}$ varies across different instances of $X_i$.
Aware that the QR factorization of $X_i$ has to be computed for each $i$,
the question is whether or not such structure is exploitable.
The answer lies in the
``Partitioned Matrix Expression'' (PME) of the QR factorization:
for a given operation,  
the PME reveals how (portions of) the output operands can be expressed in
terms of (portions of) the input operands~\cite{Cl1ck-PMEs}. 

In this specific case, we wonder if and how the vertical
partitioning of $X$ propagates
to the computation of $Q$ and $R$. Consider
\begin{equation}
    \myFlaOneByTwo{Q_L}{Q_{R}} 
    \myFlaTwoByTwo{R_{TL}}{R_{TR}}{0}{R_{BR}} =
    \myFlaOneByTwo{X_L}{X_{R}},
    \label{eq:PMEqr}
\end{equation}
where 
$
	Q_L	\in R^{n \times l}, 		
	Q_R	\in R^{n \times r}, 		
	R_{TL}	\in R^{l \times l}, 		
	R_{TR}	\in R^{l \times r},\ \text{and}\ 		
	R_{BR}	\in R^{r \times r}.
$
By rewriting Eq.~\eqref{eq:PMEqr} as
$$
    \myFlaOneByTwo{Q_L R_{TL} = X_L}{Q_L R_{TR} + Q_R R_{BR} = X_{R}},
$$
and using the orthogonality of $Q$ ($Q_L^T Q_R = 0$), 
one derives the assignments
    \begin{align}
        \{Q_L, R_{TL}\} & := qr(X_L) \nonumber \\
        R_{TR} & := Q_L^T X_{R} \label{eq:part-qr} \\
        \{Q_{R}, R_{BR}\} & := qr(X_{R} - Q_L R_{TR}) \nonumber,
    \end{align}
which indicate that the factorization of $X_L$ can be computed only once and
re-used as $X_R$ varies.

In Alg.~\ref{alg:x-exposed} all the observations made so
far are incorporated. In particular, the loop ordering is set to $i, j$, and 
code motion is applied whenever possible, to avoid redundant calculations.
Each line of the algorithm is annotated with the corresponding BLAS or LAPACK
kernel and its computational cost. 
\begin{algorithm}
    \small
\algsetup{indent=2em}
    \begin{algorithmic}[1]
    \STATE \parbox{18em}{$\{ Q_L , R_{TL} \} := qr(X_L)$}
        \parbox{1.2cm}{({\qr{}})} \parbox{1cm}{$2nl^2$}
    \FOR{i := 1 \TO m}
        \STATE {\parbox{16em}{$R_{TR_i} := Q_L^T X_{R_i}$}}
            {\parbox{1.2cm}{{(\gemm{})}}} \parbox{1cm}{$2m\mathbf{lnr}$}
        \STATE {\parbox{16em}{$T_i := X_{R_i}-Q_L R_{TR_i}$}}
            {\parbox{1.2cm}{{(\gemm{})}}} \parbox{1cm}{$2m\mathbf{lnr}$}
        \STATE {\parbox{16em}{$\{Q_{R_i}, R_{BR_i}\} := qr(T_i)$}}
            {\parbox{1.2cm}{{(\qr{})}}} \parbox{1cm}{$2m\mathbf{nr^2}$}
        \FOR{j := 1 \TO t}
            \STATE \parbox{14em}{$z_{ij} : = \myFlaTwoByOne{Q_L^T}{Q_{R_i}^T} y_j$}
                \parbox{1.2cm}{(\gemv{})} \parbox{1cm}{$2mt\mathbf{pn}$}
            \STATE \parbox{14em}{$b_j := \myFlaTwoByTwo{R_{TL}}{R_{TR_i}}{0}{R_{BR_i}}^{-1} z_{ij}$}
                \parbox{1.2cm}{(\trsv{})} \parbox{1cm}{$mt\mathbf{p^2}$}
        \ENDFOR
    \ENDFOR
    \end{algorithmic}
\caption{\bf: Structure of $X$ exposed and exploited}
\label{alg:x-exposed}
\end{algorithm}
Realizing that not only $X_L$, but also 
$Q_L$ and $R_{TL}$ are constant (they only depend on $X_L$),
we can now deliver a more sophisticated algorithm which 
saves even more flops.


As a direct effect of the partitioning of $Q_i$ in $(Q_L | Q_{R_i})$ in line 7,
the vector $z_{ij}$ 
can be decomposed as
$$\myFlaTwoByOne{z_{T_{j}}}{z_{B_{ij}}} := \myFlaTwoByOne{Q_L^T y_j}{Q_{R_i}^T y_j},$$
suggesting that the top part ($z_{T_j} := Q^T_L y_j$) may be
precomputed, once per vector $y_j$, and then reused.

Similarly, the structure of $R_i$ can be exploited to avoid redundant computation
within the triangular system in line 8. By using the same top-bottom splitting for 
$z_{ij}$, and partitioning $b_j$ accordingly, we obtain the expression
$$
    \myFlaTwoByTwo{R_{TL}}{R_{TR_i}}{0}{R_{BR_i}} 
    \myFlaTwoByOne{b_T}{b_B}  =
    \myFlaTwoByOne{z_{T_j}}{z_{B_{ij}}},
$$
which can be rewritten as
$$
\myFlaTwoByOne{R_{TL} b_{T} + R_{TR_i} b_B = z_{T_j}}{R_{BR_i} b_B = z_{B_{ij}}}.
$$
Straightforward manipulation suffices to show that $b_T$ and $b_B$ can
be computed as
\begin{align}
    b_{B_{ij}} &:= R_{BR_i}^{-1} z_{B_{ij}} \nonumber \\
    b_{T_{ij}} &:=  R_{TL}^{-1} ( z_{T_j} - R_{TR_i} b_{B_{ij}} ). \label{eq:part-trsm} 
\end{align}

Given the dependencies with the loop indices, the top part of $b$ can be
partially precomputed by first moving the operation $k_j := R_{TL}^{-1} z_{T_{ij}}$
to the same initial loop where $z_{T_j} := Q^T_L y_j$ is precomputed,
and then moving the operation $h_i := R_{TL}^{-1} R_{TR_i}$ out of the innermost loop.
The resulting algorithm is displayed in Alg.~\ref{alg:qr-exposed}.
As discussed, $z_{T_j}$ and $k_j$ are precomputed in an initial loop (lines
2--5), and $k_j$ is kept in memory (a few megabytes at most)
for later use within the nested double-loop (line 14). Also, the computation of $h_i$ is
taken out of the innermost loop (line 10) and reused within the
innermost loop (line 14). The cost of Alg.~\ref{alg:qr-exposed} 
is dominated by the term $2mtrn$, 
that is, a factor of ${p}/{r}$ fewer operations 
with respect to Alg.~\ref{alg:x-exposed}.

\begin{algorithm}
    \small
   \algsetup{indent=2em}
     \begin{algorithmic}[1]
         \STATE \parbox{16em}{$\{Q_L , R_{TL}\} := qr(X_L)$}\parbox{1.5cm}{(\qr{})} \parbox{1cm}{$2nl^2$}
 	\FOR{j := 1 \TO t} 
        \STATE {\parbox{14em}{$z_{T_j} :=	Q_{L}^T y_j$}}{\parbox{1.5cm}{(\gemv{})}} \parbox{1cm}{$2t\mathbf{ln}$}
        \STATE {\parbox{14em}{$k_j := R_{TL}^{-1} z_{T_j}$}}{\parbox{1.5cm}{(\trsv{})}} \parbox{1cm}{$t\mathbf{l^2}$}
	\ENDFOR
	\FOR{i := 1 \TO m}
        \STATE {\parbox{14em}{$R_{TR_i} := Q_L^T X_{R_i}$}}{\parbox{1.5cm}{{(\gemm{})}}} \parbox{1cm}{$2m\mathbf{lnr}$}
        \STATE {\parbox{14em}{$T_i := X_{R_i}-Q_L R_{TR_i}$}}{\parbox{1.5cm}{{(\gemm{})}}} \parbox{1cm}{$2m\mathbf{lnr}$}
        \STATE {\parbox{14em}{$\{Q_{R_i} , R_{BR_i}\} := qr(T_i)$}}{\parbox{1.5cm}{{(\qr{})}}} \parbox{1cm}{$2m\mathbf{nr^2}$}
        \STATE {\parbox{14em}{$h_i := R_{TL}^{-1} R_{TR_i}$}}{\parbox{1.5cm}{{(\trsm{})}}} \parbox{1cm}{$m\mathbf{l^2r}$}
      	\FOR{j := 1 \TO t}
            \STATE \parbox{12em}{$z_{B_{ij}} := Q_{R_i}^T y_j$}\parbox{1.5cm}{(\gemv{})} \parbox{1cm}{$2mt\mathbf{rn}$}
            \STATE \parbox{12em}{$b_{B_{ij}} :=  R_{BR_i}^{-1} z_{B_{ij}}$}\parbox{1.5cm}{(\trsv{})} \parbox{1cm}{$mt\mathbf{r^2}$}
            \STATE \parbox{12em}{$b_{T_{ij}} :=  k_j - h_i b_{B_{ij}}$}\parbox{1.5cm}{(\gemv{})} \parbox{1cm}{$2mt\mathbf{lr}$}
  		\ENDFOR
  \ENDFOR
  \end{algorithmic}
\caption{\bf: Structure of $Q$ and $R$ exposed and exploited}
\label{alg:qr-exposed}
\end{algorithm}

\section{Out-of-core algorithm: Analysis of data streaming}
\label{sec:ooc}

As mentioned in Sec.~\ref{sec:intro},
the second challenge to be addressed when dealing with series and grids of
least squares problems is the management of large datasets.
An example clarifies the situation; in the 
characteristic scenario in which $n=30{,}000$, $p=10$ ($r=2$), 
$m=10^7$, and $t=10^4$, the input
dataset is of size 2.4 TBs, and the computation generates 4 TBs as output.
Since such datasets exceed the main memory capacity of current
shared-memory nodes and therefore reside on disk, one has to resort to an
out-of-core algorithm~\cite{Toledo-OOC-survey-99}.
The main idea behind our design is to effectively {\em tile} (block) the
computation to amortize the time spent in moving data. 

To emphasize the need for tiling, we commence by discussing the I/O requirements
of a naive out-of-core implementation of Alg.~\ref{alg:qr-exposed}, as
sketched in Alg.~\ref{alg:ooc-naive}.
First, the algorithm requires the loading of the entire set of vectors $y_j$
(line 2) for the computations in the initial loop. Then, 
it loads once the entire set of matrices $X_{R_i}$ (line 6), 
loads $m$ times the entire set of vectors $y_j$ (line 9), and 
finally requires the storage of the resulting $m \times t$ vectors $b_{ij}$ (line 11).
The reading of $y_j$ for a total of $m$ times (line 9) is the clear
I/O bottleneck. For the aforementioned example, the algorithm generates
12 petabytes of disk-to-memory traffic, which at a (rather optimistic) transfer
rate of 2 GBytes/sec, would take 70 days of I/O. 
It is thus imperative to reorganize I/O and computation to greatly reduce
the amount of generated I/O traffic.

\begin{algorithm}
    \small
   \algsetup{indent=2em}
     \begin{algorithmic}[1]
    \STATE {\parbox{16em}{\text{\tt Load matrix} $X_L$}\parbox{5cm}{$\quad 4 l n \rightarrow$ $<$ 1 MBs} }
	\FOR{j := 1 \TO t} 
        \STATE {\parbox{14em}{\text{\tt Load vector} $y_j$}\parbox{5cm}{$\quad 4 t n \rightarrow$ 1.2 GBs} }
		\STATE {\parbox{14em}{\tt Compute with $y_j$}}
	\ENDFOR
	\FOR{i := 1 \TO m}
        \STATE {\parbox{15em}{\tt Load matrix $X_{R_i}$}\parbox{5cm}{$4 m n r \rightarrow$ 2.4 TBs} }
        \STATE {\parbox{14em}{\tt Compute with $X_{R_i}$}}
        \FOR{j := 1 \TO t}
            \STATE {\parbox{12em}{\tt Load vector $y_j$}\parbox{5cm}{$\quad 4 m t n \rightarrow$ 12 PBs} }
            \STATE {\parbox{12em}{\tt Compute with $X_{R_i}, y_j$}}
            \STATE {\parbox{12em}{\tt Store vector ${\beta}_{ij}$}\parbox{5cm}{$\quad 4 m t p \rightarrow$ 4 TBs} }
        \ENDFOR
	\ENDFOR
\end{algorithmic}
\caption{{\bf: Naive out-of-core approach}}
\label{alg:ooc-naive}
\end{algorithm}

\subsection{Analysis of computational cost over data movement}

A {\em tiled} algorithm decomposes the computation 
of the 2D grid of problems into subgrids or {\em tiles}.
Instead of loading one single matrix $X_{R_i}$ and vector $y_j$,
and storing one single vector $b_{ij}$, the idea is to load and store
multiple of them in a {\it slab}. 
The tiled algorithm presented in Alg.~\ref{alg:ooc-tiled}
loads slabs of $t_b$ vectors $y_j$ ($\hat{Y}$) and
$m_b$ matrices $X_{R_i}$ ($\hat{X}_R$) in lines 4, 11, and 19; and
stores slabs of computed $m_b \times t_b$ vectors $b_{ij}$ ($\hat{B}$)
in line 27.

\begin{algorithm}
    \small
   \algsetup{indent=1em}
     \begin{algorithmic}[1]
    \STATE {\parbox{19em}{\textcolor{io}{\text{\tt Load} $X_L$}}}
    \STATE \parbox{17em}{$\{Q_L, R_{TL}\} := qr(X_L)$}\parbox{1.2cm}{(\qr{})} \parbox{1cm}{$2nl^2$}
    \FOR{j := 1 \TO $t/t_b$} 
        \STATE {\parbox{17em}{\textcolor{io}{\text{\tt Load slab} $\hat{Y}_j$}}} 
        \FOR{l := 1 \TO $t_b$} 
        \STATE {\parbox{15em}{$\hat{Z}_{T_{j_l}} := Q_{L}^T \hat{Y}_{j_l}$}}{\parbox{1.2cm}{(\gemv{})}} \parbox{1cm}{$2t\mathbf{ln}$} 
        \STATE {\parbox{15em}{$\hat{K}_{j_l} := R_{TL}^{-1} \hat{Z}_{T_{j_l}}$}}{\parbox{1.2cm}{(\trsv{})}} \parbox{1cm}{$t\mathbf{l^2}$}
        \ENDFOR
    \ENDFOR
    \FOR{i := 1 \TO $m/m_b$}
        \STATE {\parbox{18em}{\textcolor{io}{\tt Load slab $\hat{X}_{R_i}$}}} 
        \FOR{k := 1 \TO  $m_b$} 
            \STATE {\parbox{15em}{$\hat{R}_{TR_{i_k}} := Q_L^T \hat{X}_{R_{i_k}}$}}{\parbox{1.2cm}{{(\gemm{})}}} \parbox{1cm}{$2m\mathbf{lnr}$}
            \STATE {\parbox{15em}{$\hat{T}_{i_k} := \hat{X}_{R_{i_k}}-Q_L \hat{R}_{TR_{i_k}}$}}{\parbox{1.2cm}{{(\gemm{})}}} \parbox{1cm}{$2m\mathbf{lnr}$}
            \STATE {\parbox{15em}{$\{\hat{Q}_{R_{i_k}}, \hat{R}_{BR_{i_k}}\} := qr(\hat{T}_{i_k})$}}{\parbox{1.2cm}{{(\qr{})}}} \parbox{1cm}{$2m\mathbf{nr^2}$}
            \STATE {\parbox{15em}{$\hat{H}_{i_k} :=  R_{TL}^{-1} \hat{R}_{TR_{i_k}}$}}{\parbox{1.2cm}{{(\trsm{})}}} \parbox{1cm}{$m\mathbf{l^2r}$}
        \ENDFOR
        \FOR{j := 1 \TO $t/t_b$}
            \STATE {\parbox{15em}{\textcolor{io}{\tt Load slab $\hat{Y}_j$}}} 
            \FOR{k := 1 \TO  $m_b$} 
                \FOR{l := 1 \TO  $t_b$} 
                    \STATE \parbox{13em}{$\hat{Z}_{B_{i_kj_l}} :=  \hat{Q}_{R_{i_k}}^T \hat{Y}_{j_l}$}\parbox{1.2cm}{(\gemv{})} \parbox{1cm}{$2mt\mathbf{rn}$}
                    \STATE \parbox{13em}{$\hat{B}_{B_{i_kj_l}} :=  R_{BR_{i_k}}^{-1} \hat{Z}_{B_{i_kj_l}}$}\parbox{1.2cm}{(\trsv{})} \parbox{1cm}{$mt\mathbf{r^2}$}
                    \STATE \parbox{13em}{$B_{T_{i_kj_l}} :=   \hat{K}_{j_l} - \hat{H}_{i_k} \hat{B}_{B_{i_kj_l}}$}\parbox{1.2cm}{(\gemv{})} \parbox{1cm}{$2mt\mathbf{lr}$}
                \ENDFOR
            \ENDFOR
            \STATE {\parbox{15em}{\textcolor{io}{\tt Store slab $\hat{B}_{ij}$}}} 
        \ENDFOR
    \ENDFOR
  \end{algorithmic}
\caption{{\bf: Tiled algorithm: Data loaded and stored in slabs}}
\label{alg:ooc-tiled}
\end{algorithm}

With tiling, the amount of I/O required by line 19 
of Alg.~\ref{alg:ooc-tiled} is constrained to
$$t \times n \times \frac{m}{m_b},$$
and, most importantly, can be adjusted by setting the 
parameter $m_b$.
We note that, in terms of I/O overhead,
there is no need to set $m_b$ to the 
largest possible value allowed by the available main memory:
it suffices to choose $m_b$ large enough so that the I/O
bottleneck shifts to the loading of $X_R$ and writing of $B$:
 $$t \times n \times \frac{m}{m_b} \ll m\times n \times r + m\times t\times p.$$

After choosing a sufficiently large value for $m_b$, 
the ratio of computation over data movement is
\begin{equation}
    \frac{O(mtrn)}{O(mnr + mtp)} \equiv \frac{O(trn)}{O(nr + tp)}.
    \label{eq:ratio}
\end{equation}
Therefore, beyond the freedom in parameterizing $m_b$, 
{\em it is the actual instance of the equation, that is, the problem sizes,
which determines whether the computation is compute-bound or IO-bound}.
We illustrate the different scenarios (memory-bound vs IO-bound) in
Sec.~\ref{sec:experiments}, where we present experimental results.

\subsection{Overlapping I/O with computation: double buffering}

For problems that are not largely dominated by I/O, it is possible
to reduce or even completely eliminate the overhead due to I/O operations
by overlapping I/O with computation.  In the development of {\sc ols-grid}, 
we adopted the well-known double buffering mechanism 
to asynchronously load and store data:
the main memory is logically split into two buffers;
while operations within an iteration of the innermost loop are performed on
the slabs previously loaded in one of the buffers, a dedicated I/O thread 
operating on the other buffer
downloads the results of the previous iteration and 
uploads the slabs for the next one. 
The experiments in Sec.~\ref{sec:experiments} confirm that this mechanism
mitigates the negative effect of data transfers and, for
compute-bound scenarios, it prevents I/O from limiting
the scalability of our solver.

\section{High performance and scalability}
\label{sec:hp}

In the previous sections, we designed an algorithm that
avoids redundant computations, and studied the impact of data transfers
to constrain (or even eliminate) the overhead due to I/O.
One last issue remains to be addressed: how to
exploit shared-memory parallelism.
In this section, we illustrate how to reorganize the computation so that it
can be cast in terms
of efficient BLAS-3 operations, and discuss how to combine
different types of parallelism to attain scalability.

\begin{figure}[!ht]
    \centering
        \includegraphics[width=0.8\textwidth]{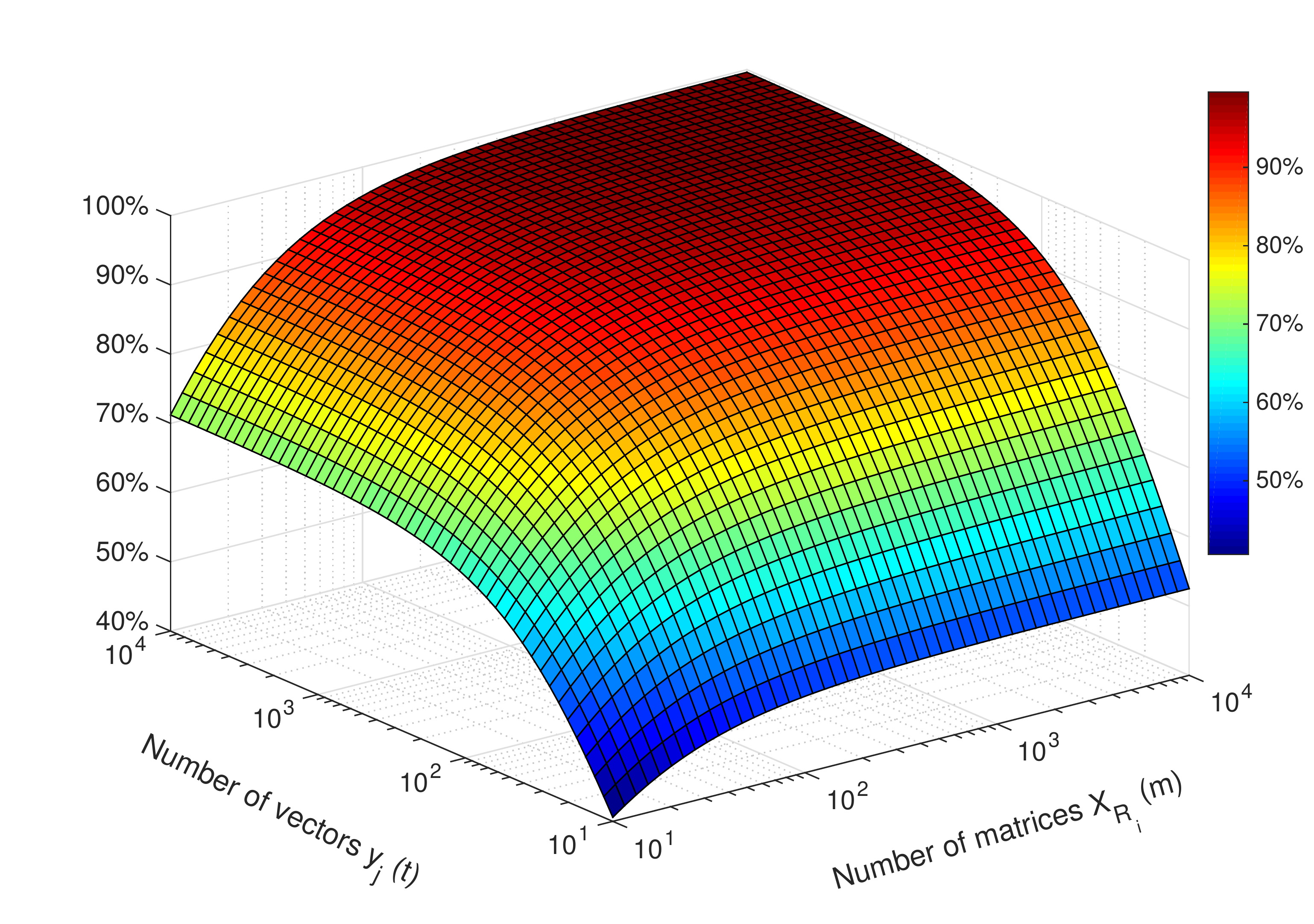}
        \caption{Percentage of the computation performed by the operation
            $Q^T_{R_i} y_j$ from the total operations required by Algorithm~\ref{alg:ooc-tiled}.}
        \label{fig:comp-bottleneck}
\end{figure}

The computational bottleneck of Alg.~\ref{alg:ooc-tiled}
is the operation $Q^T_{R_i} y_j$ in line 22,
which is executed $m \times t\,$ times.
This is confirmed visually by
Fig.~\ref{fig:comp-bottleneck},
which presents,
for $n = 1{,}000$ and varying $m$ and $t$, 
the weight of that operation as a percent of the total computation (flops) 
performed by the algorithm.
Already for small values of $m$ and $t$ (in the hundreds), the operation
accounts for 90\% of the computation. 
When $m$ and $t$ take more realistic values (in the thousands
or more), the percent attains values close to 100\%.

The operation $Q^T_{R_i} y_j$ is implemented by
the inefficient BLAS-2 {\sc gemv} kernel, which on
a single-core only attains about 10\% of the peak performance, and on
multi-cores suffers from poor scalability. 
The idea to overcome this bottleneck 
is to combine the $m_b$
matrices $Q^T_{R_i}$ into a block $\hat{Q}^T_{R_i}$, and the $t_b$ vectors $y_j$
into a block $\hat{Y}_j$.
This transformation effectively recasts the small matrix-vector 
products ({\sc gemv}'s) into the
large and much more efficient matrix-matrix operation
$\hat{Q}^T_{R_i} \hat{Y}_j$ ({\sc gemm}), which achieves 
both close-to-optimal performance and high scalability.

While the use of a multi-threaded version of the BLAS library
for the \gemm{} operation $\hat{Q}^T_{R_i} \hat{Y}_j$ enables
high scalability, it is a poor choice for other sections
of the algorithm, which do not scale as nicely. In fact, when
using a large number of cores (as is the case
in our experiments), the weight of these other sections increases
to the point that it affects the overall scalability.
As a solution to mitigate this problem we turn to a hybrid
parallelism combining multi-threaded BLAS with OpenMP.
Lines 13, 14, and 16 in Alg.~\ref{alg:ooc-tiled} are examples of operations
that do not scale with a mere use of a multi-threaded BLAS.
Let us illustrate the issue with the matrix product in line 13.
One of these \gemm{}'s in isolation multiplies $Q^T_L \in R^{l \times n}$
with $X_{R_i} \in R^{n \times r}$; that is, it multiplies a wide matrix with only a few 
rows times a thin matrix with only a few columns. Even when $m_b$ $X_{R_i}$'s
are combined into the block $\hat{X}_{R_i}$, one of the matrices ($Q_L$) is
still rather small. In terms of number of operations per element, the ratio is
very low, which results in an inefficient and poorly scalable operation.
In this case, using a multi-threaded BLAS is not sufficient (in our
experiments we observed a performance of about 5 GF/s, i.e., below 1\% efficiency). 
Instead, we decide to
explicitly split the operation among the compute threads by means
of OpenMP directives. Each thread takes a proportional number of columns
from $\hat{X}_{R_i}$ and computes the corresponding \gemm{} using a single-threaded
BLAS call. This second alternative results in a speedup of 5x to 10x,
sufficient to mitigate the impact in the overall performance and
scalability.

The resulting algorithm is displayed in Alg.~\ref{alg:final}. Among the
computational bottlenecks, we highlight in green the operations
parallelized by means of a multi-threaded version of BLAS (line 18)
and in blue the operations parallelized using OpenMP (lines 10--12).
As we show in the next section, the recasting of small operations into larger
ones and the use of a hybrid form of parallelism
that combines multi-threaded BLAS and OpenMP
leads to satisfactory performance and scalability
signatures.

\begin{algorithm}
    \small
   \algsetup{indent=1em}
     \begin{algorithmic}[1]
    \STATE {\parbox{19em}{\textcolor{io}{\text{\tt Load} $X_L$}}}
    \STATE \parbox{16em}{$\{Q_L, R_{TL}\} := qr(X_L)$}\parbox{1.2cm}{(\qr{})} \parbox{1cm}{$2nl^2$}
 	\FOR{j := 1 \TO $t/tb$} 
        \STATE {\parbox{17em}{\textcolor{io}{\text{\tt Load slab} $\hat{Y}_j$}}} 
        \STATE {\parbox{15em}{$\hat{Z}_{T_j} := Q_{L}^T \hat{Y}_j$}}{\parbox{1.2cm}{(\gemm{})}} \parbox{1cm}{$2t\mathbf{ln}$} 
        \STATE {\parbox{15em}{$ \hat{K}_j := R_{TL}^{-1} \hat{Z}_{T_j}$}}{\parbox{1.2cm}{(\trsm{})}} \parbox{1cm}{$t\mathbf{l^2}$}
	\ENDFOR
	\FOR{i := 1 \TO $m/mb$}
        \STATE {\parbox{18em}{\textcolor{io}{\tt Load slab $\hat{X}_{R_i}$}}} 
        \STATE {\parbox{15em}{\textcolor{openmp}{$\hat{R}_{TR_i} := 	Q_L^T \hat{X}_{R_i}$}}}{\parbox{1.2cm}{{(\gemm{})}}} \parbox{1cm}{$2m\mathbf{lnr}$}
        \STATE {\parbox{15em}{\textcolor{openmp}{$\hat{T}_i :=  \hat{X}_{R_i}-Q_L \hat{R}_{TR_i}	$}}}{\parbox{1.2cm}{{(\gemm{})}}} \parbox{1cm}{$2m\mathbf{lnr}$}
        \STATE {\parbox{15em}{\textcolor{openmp}{$\hat{H}_i :=  R_{TL}^{-1} \hat{R}_{TR_i}$}}}{\parbox{1.2cm}{{(\trsm{})}}} \parbox{1cm}{$m\mathbf{l^2r}$}
		\FOR{k := 1 \TO  $m_b$} 
            \STATE {\parbox{14em}{$\{\hat{Q}_{R_{i_k}}, \hat{R}_{BR_{i_k}}\} := qr(\hat{T}_{i_k})$}}{\parbox{1.2cm}{{(\qr{})}}} \parbox{1cm}{$2m\mathbf{nr^2}$}
		\ENDFOR
        \FOR{j := 1 \TO $t/tb$}
            \STATE {\parbox{15em}{\textcolor{io}{\tt Load slab $\hat{Y}_j$}}} 
            \STATE \parbox{14em}{\textcolor{mtblas}{$\hat{Z}_{B_{ij}} :=  \hat{Q}_{R_i}^T \hat{Y}_j$}}\parbox{1.2cm}{(\gemm{})} \parbox{1cm}{$2mt\mathbf{rn}$}
  			\FOR{k := 1 \TO  $m_b$} 
            \STATE \parbox{13em}{$\hat{B}_{B_{i_kj}} :=  R_{BR_{i_k}}^{-1} \hat{Z}_{B_{i_kj}}$}\parbox{1.2cm}{(\trsm{})} \parbox{1cm}{$mt\mathbf{r^2}$}
            \STATE \parbox{13em}{$B_{T_{i_kj}} :=   \hat{K}_j - \hat{H}_{i_k} \hat{B}_{B_{i_kj}}$}\parbox{1.2cm}{(\gemm{})} \parbox{1cm}{$2mt\mathbf{lr}$}
			\ENDFOR
            \STATE {\parbox{15em}{\textcolor{io}{\tt Store slab $\hat{B}_{ij}$}}} 
  		\ENDFOR
  \ENDFOR
  \end{algorithmic}
  \caption{{\bf: Efficient and scalable algorithm: \omicols{}}}
\label{alg:final}
\end{algorithm}

\section{Experimental results}
\label{sec:experiments}

In this section, {\sc ols-grid} (the implementation of Alg.~\ref{alg:final})
is tested in a variety of settings to provide
evidence that it makes a nearly optimal use of the available resources;
it is also compared and contrasted with a number of available
alternatives, namely 
a naive solver (the linear regression solver from ProbABEL), 
a specialized solver for grids of GLS (generalized least squares) problems (OmicABEL-Eig), and 
a specialized solver for grids of OLS problems (GWASP).

All our tests were run on a system consisting of 4 Intel(R) Xeon(R) E7-4850
Westmere-EX multi-core processors. Each processor comprises 10 cores operating
at a frequency of 2.00 GHz, for a combined peak performance in single
precision of 640~GFlops/sec.
The system is equipped with 256~GBs of RAM and 8~TBs of disk as secondary
memory; our measurements indicate that the Lustre file system attains 
a maximum bandwidth of about 300~MBs/sec and 1.7~GBs/sec for
writing and reading operations, respectively.
The solver was compiled using Intel's icc compiler (v14.0.1), and
linked to Intel's MKL multi-threaded library (v14.0.1), which provides
for BLAS and LAPACK functionality. The routine makes use of the OpenMP
parallelism provided by the compiler through a number of {\it pragma}
directives. All computations were performed in single precision,
and the correctness of the results was assessed by direct comparison with
the OLS solver from LAPACK ({\sc sgels}).

For the asynchronous I/O transfers, we initially tested the AIO library 
(available in all Unix systems). We observed that I/O operations had
a considerable impact in performance by limiting the flops attained by
\gemm{}. Since AIO does not offer the means to specify the amount of threads
spawned for I/O and to pin threads to cores, we developed
our own light-weight library which uses one single thread for I/O operations and
allows thread pinning.\footnote{The library is publicly available at
\url{http://hpac.rwth-aachen.de/~fabregat/software/lwaio-v0.1.tgz}.}
  
\subsection{Compute-bound vs IO-bound scenarios}

\begin{center}
\begin{figure}[!ht]
  \centering
    \includegraphics[width=0.8\textwidth]{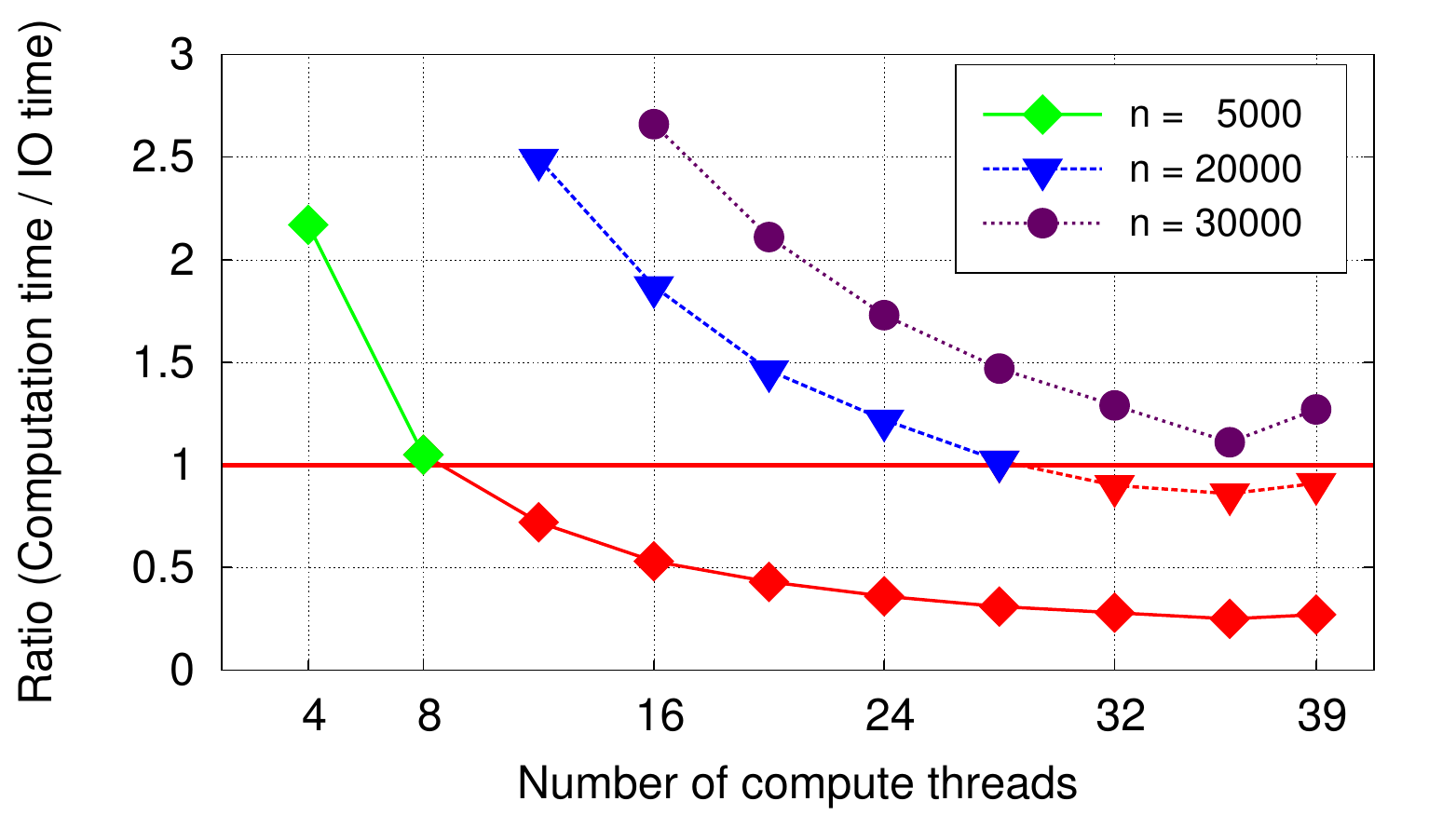}
  \caption{Ratio of compute time over I/O time. Results for a varying value
  of $n$ and an increasing number of compute threads. The other problem
  sizes are fixed: $l=4$, $r=1$, $m=10^6$, $t=10^4$, $m_b=10^4$, and $t_b=5\times10^3$.}
  \label{fig:IOvsComp}
\end{figure}
\end{center}

We commence the study of our own solver by showing the practical implications of the
ratio computation over I/O transfers
$$
    \frac{O(trn)}{O(nr + tp)}.
$$
We collected the time spent in computations and the time
spent in I/O operations for a range of problems varying
the size of $n$.
Figure~\ref{fig:IOvsComp} presents these results. 
The horizontal red line corresponds to a ratio of 1, 
while the other three lines represent the ratio compute time
over I/O time for different values of $n$ and for an increasing number of compute threads.

While the bandwidth is fixed, the computational power increases
with the number of compute threads.
For $n=5{,}000$ and $n=20{,}000$, the I/O time eventually 
overtakes computation time (the ratio becomes $<$ 1). 
At that point, the problem becomes
IO-bound and the use of additional compute threads does not
reduce the time-to-solution. 
For $n=30{,}000$ instead, the time spent in I/O never
grows larger than the time spent in computation and I/O is perfectly overlapped.

In the next experiment we show how the results in Fig.~\ref{fig:IOvsComp}
translate into scalability results.

\subsection{Scalability}

\begin{figure}[!ht]
  \centering
  \includegraphics[width=.8\textwidth]{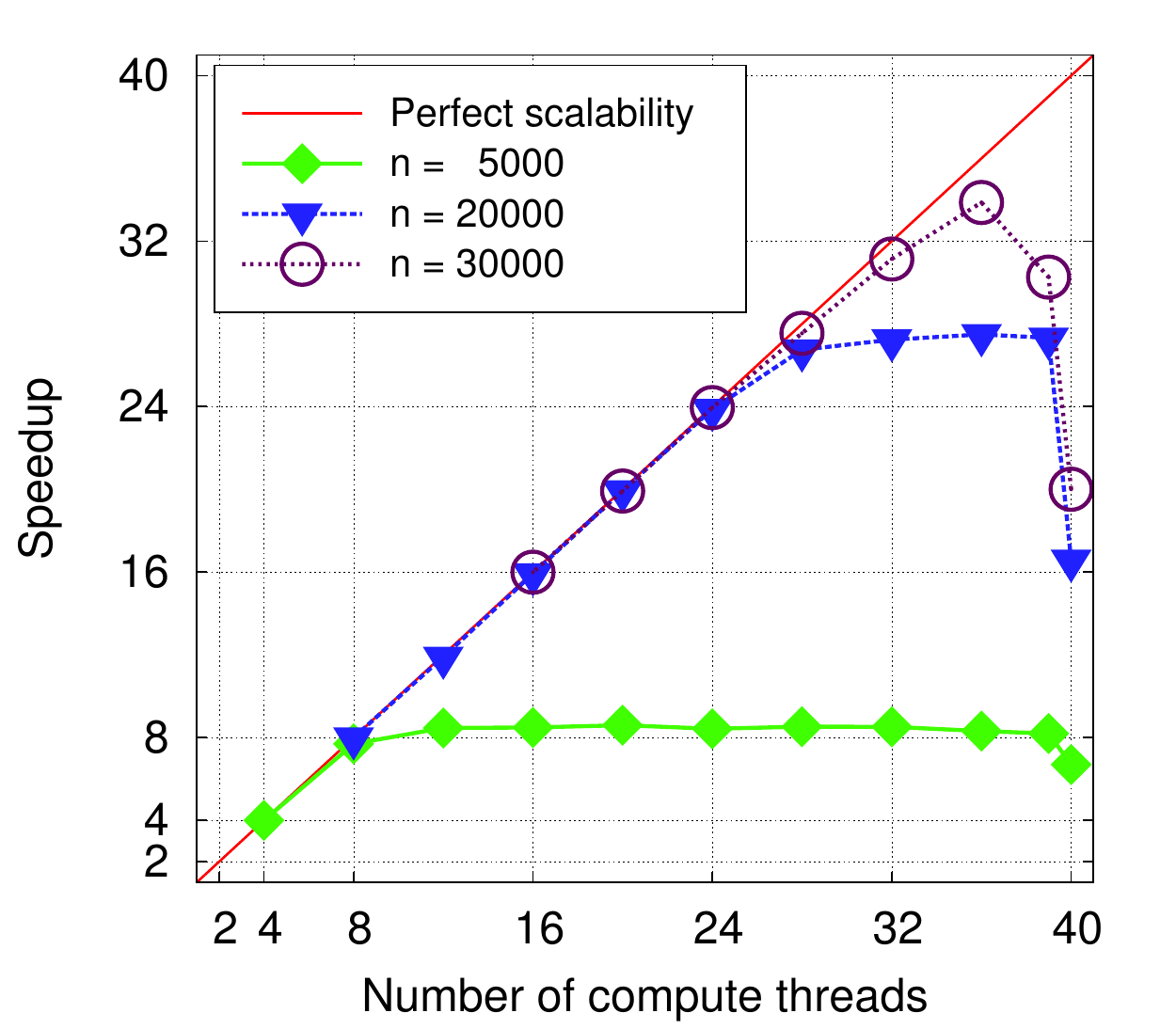}
  \caption{Scalability results for a varying value of $n$. The other problem
  sizes are fixed: $l=4$, $r=1$, $m=10^6$, $t=10^4$, $m_b=10^4$, and $t_b=5\times10^3$.}
    \label{fig:scal}
\end{figure}

We study now the scalability of {\sc ols-grid}. 
Figure~\ref{fig:scal} presents scalability results 
for the same problem sizes discussed in the previous experiment.
The red diagonal line represents perfect scalability.
As the ratio in Fig.~\ref{fig:IOvsComp} suggested,
the line for $n=5{,}000$ shows perfect scalability for
up to 8 threads; from that point on, the I/O transfers dominate the
execution time and no larger speedups are possible.
Similarly, for $n=20{,}000$, almost perfect
scalability is attained with up to 28 compute threads. 
Again, beyond
that number, the I/O dominates and the scalability plateaus.
Instead, for $n=30{,}000$, the I/O never dominates
execution time and the solver attains speedups of around 34x
when 36 compute threads are used.
In all cases, we observe a drop in scalability when
40 compute threads are used; this is because one compute thread 
and the I/O thread share one core.
We attribute the drop at 39 compute threads for
the experiment with $n=30{,}000$---and, less apparently,
at $n=5{,}000$ and $n=20{,}000$---to potential
memory conflicts, since in the used architecture
each pair of threads share the L1 and L2 caches.

An important message to extract from these results is the need to
understand the characteristics of the problem at hand to decide
which architecture fits best our needs. In the case of omics
GWAA, the size of $n$ plays an important role in the decision
of whether investing in further computational power or larger
bandwidth.

\subsection{Efficiency}

\begin{figure}[!ht]
  \centering
      \includegraphics[width=1.0\textwidth]{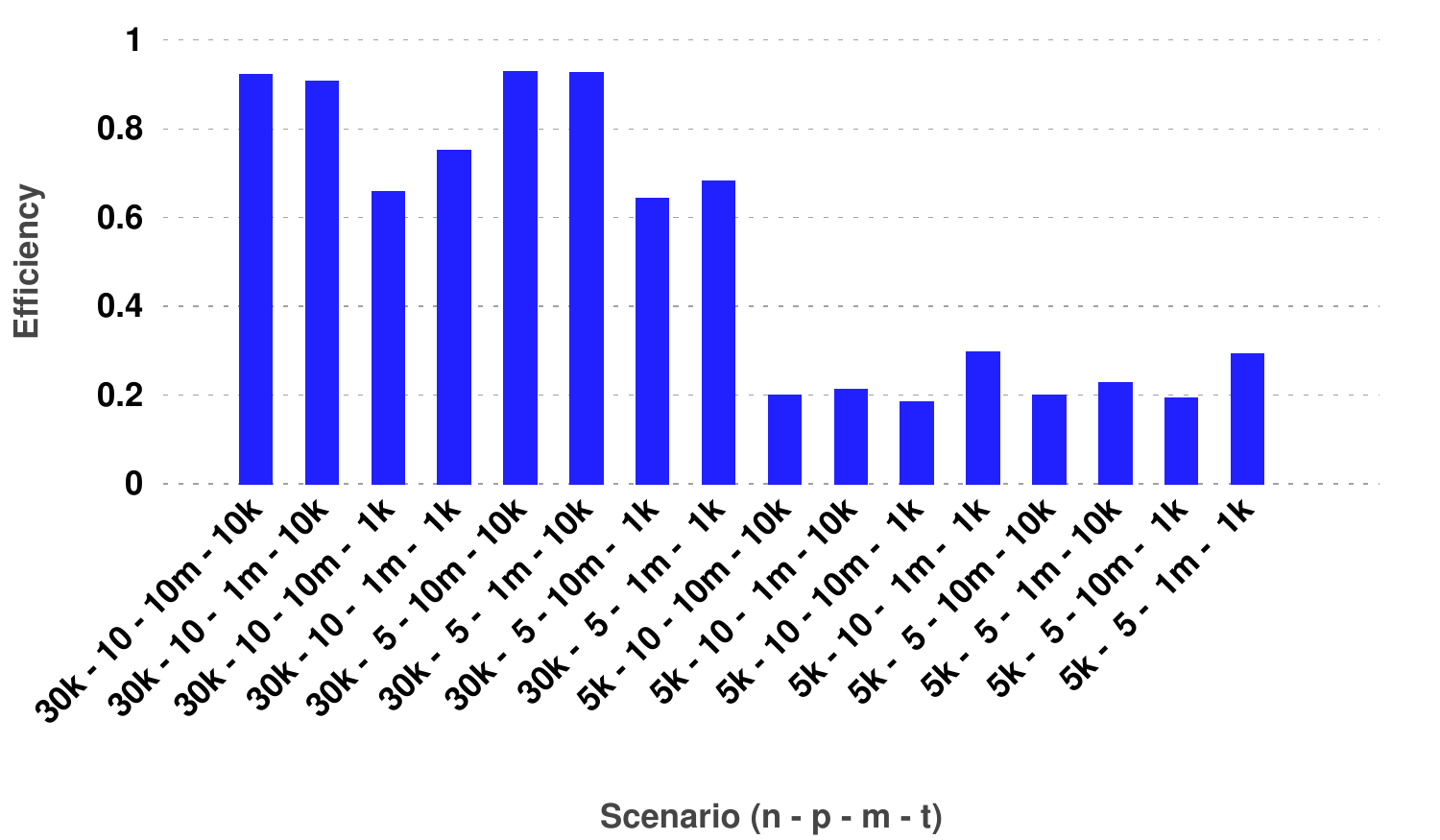}
  \caption{Efficiency of our solver for a range of scenarios
      using 36 compute threads. \gemm{}'s peak performance
      is used as practical peak performance of the architecture.}
	\label{fig:part_percent}
\end{figure}

As a final result, we quantify how efficiently {\sc ols-grid} uses
the available resources. To this end, we measured the time-to-solution
for a variety of scenarios using 36 compute threads and compared
the performance with that of the practical peak performance, that is,
the best performance attained by \gemm{}.
The results are presented in Fig.~\ref{fig:part_percent};
the tested scenarios include all combinations of $n=(5{,}000, \, 30{,}000)$,
$p=(5, 10)$, $m=(10^6, 10^7)$, and $t=(10^3, 10^4)$.
While for small sizes of $n$ the I/O transfers limit the efficiency,
with a maximum of 0.3, for large values of $n$ the attained efficiency 
ranges from 0.64 to 0.93.

\subsection{Comparison with alternative solvers}
\label{sec:comparison}

We conclude the performance study with a detailed discussion
on how {\sc ols-grid} compares to other existing solvers.
All presented timings were obtained in the same experimental setup 
as the previous sections, and making use of the available 40 cores.

\subsubsection{ProbABEL}

is a collection of tools for linear and logistic regression,
as well as Cox proportional hazards models~\cite{Aulchenko-Struchalin-2010}.
While in this discussion
we concentrate on its deficiencies in terms of performance
for large-scale linear regression analyses, we want to clarify that
this is a versatile tool offering functionality well beyond the core linear
regression solver.
ProbABEL's OLS solver does not take advantage
of the structure of the problem at hand, and thus delivers poor performance.
For instance, an analysis with sizes $n=30{,}000$, $p=5$ ($l=4$, $p=1$),
$m=10{,}000$, and $t=100$ takes almost half an hour, 
while {\sc ols-grid} completes in less than two seconds. 
The gap in performance between a traditional black-box solver
and a carefully designed solver is enormous: 
if the previous analysis were extended to $m=10^7$ and $t=10^4$, 
ProbABEL's algorithm would become unusable 
(estimated completion time of more than 13 years), 
while our {\sc ols-grid} completes the analysis in 6.9 hours. 

\subsubsection{OmicABEL-Eig} is the multi-trait ($t > 1$) solver for
GLS-based analyses in OmicABEL~\cite{OmicABEL,MultiGWAS}.
The concepts behind OmicABEL-Eig are
similar to those discussed in this paper; in particular, the algorithm 
relies on the reduction of GLS problems to OLS ones. However,
the dependencies in the grid to be solved are different, 
and computationally the consequences are huge. 
On the one hand, while the asymptotic complexity of both OmicABEL-Eig
and \omicols{} is the same, the constants for the former are large. On the other hand,
even though OmicABEL-Eig shows great scalability, its efficiency compared to
the present solver is low because it cannot take advantage of \gemm{}.
As a result, the same analysis that took \omicols{} 6.9 hours to complete,
would require a month of computation for OmicABEL-Eig.
In short, even the use of an existing fully optimized solver for a similar
regression analysis may not be the best choice, and it is worth the
effort of designing a new one tailored to the needs of the specific problem at hand.

\subsubsection{GWASP} is a recently developed solver tailored for the computation
of a grid of OLS problems similar to that addressed in this paper~\cite{gwasp}.
Thanks to techniques such as 
the aggregation of vector-vector and matrix-vector operations into
matrix-vector and matrix-matrix operations, respectively, 
GWASP delivers an efficient algorithm.
Unfortunately, a direct comparison with {\sc ols-grid} is not possible, 
since
1) this solver only computes the last $r$ entries of $\beta_{ij}$,
2) the focus is on single-trait ($t = 1$) analyses, and
3) only R code is provided.

In order to study GWASP's performance, we implemented the
algorithm in C using the BLAS and LAPACK libraries, and ran a number of
single-trait experiments.
For a problem of size $n=5{,}000$, $p=5$ 
($l=4$, $r=1$), $m=10^6$, and $t=1$, GWASP took 61 seconds, while
\omicols{} completed in 14 seconds. 
Similarly, if $n$ is increased to $n=30{,}000$, the analysis completes
in 314 and 84 seconds for GWASP and \omicols{}, respectively.
Since 
the computational complexity of both solvers is similar and
these scenarios are IO-bound,
the differences mainly
reside in the careful hybrid parallelization of \omicols{}
and the overlapping of data transfers.

For the general case of $t>1$, 
GWASP may only be used one trait at
a time. In the last scenario with $t=1{,}000$, it would require days
to complete. By contrast, \omicols{} finishes in 138 seconds,
thanks to the optimizations enabled when considering the 
2D grid of problems as a whole.
Despite this relatively large gap, we believe that GWASP can be extended
to multi-trait analyses and to the computation of entire $\beta$'s, while
achieving performance comparable to that of our solver.

\section{Conclusions}
\label{sec:conclusions}

We addressed the design and implementation of efficient solvers for large-scale
linear regression analyses. As case study, we focused on the computation of a
two-dimensional grid of ordinary least squares problems as it appears in the
context of genome-wide association analyses.
The resulting routine, \omicols{}, showed to be highly efficient and
scalable. 

Starting from
the mathematical description of the problem, we designed an
incore algorithm that exploits the available problem-specific knowledge and structure.
Next, to enable the solution of problems with large datasets that do
not fit in today multi-core's main memory, we transformed the incore
algorithm into an out-of-core one, which, thanks to tiling,
constrains the amount of required
data movement. By incorporating the double-buffering technique and
using an asynchronous I/O library, we completely eliminated I/O overhead
whenever possible.

Finally, by reorganizing the calculations, the computational bottleneck
was cast in terms of the highly efficient \gemm{} routine. Combining
multi-threaded BLAS and OpenMP parallelism, \omicols{}
also attains high performance and high scalability. More specifically,
for large enough analyses, the solver achieves single-core efficiency 
beyond 90\% of peak performance, and speedups of up to 34x with
36 cores.

While previously existing tools allow for multiple trait analysis,
these are limited to small datasets and require considerable runtimes.
We enable the analysis of previously intractable datasets and offer shorter 
times to solution. Our \omicols{} solver is already integrated in the 
GenABEL suite, and adopted by a number of research groups.

\subsection{Future work}

Two main research directions remain open. On the one hand,  
support for distributed-memory architectures is desirable, 
allowing for further reduction in time-to-solution. 
This step will require a careful distribution of
workload among nodes and the use of advanced I/O techniques to prevent 
data movement from becoming a bottleneck.

On the other hand, we are already working on the support for so-called {\em
  missing} and {\em erroneous data}.
Often, large datasets are collected by combining data from different
sources. Therefore, subsets of data for certain individuals (for instance,
entries in $y_j$ vectors) may not be available and labeled as missing values (NaNs).
Erroneous data may origin, for instance, from errors in the acquisition.
Accepting input data with NaN values (and preprocessing it accordingly)
will make our software of broader appeal
and will facilitate its adoption by the broad GWAA community.

\subsection*{\bf Acknowledgments}

Financial support from the Deutsche Forschungsgemeinschaft (German Research
Association) through grant GSC 111 is gratefully acknowledged. The authors
thank Yurii Aulchenko for fruitful discussions on the biological background of
GWAA.


\end{document}